\documentclass[preprint,prl]{revtex4}

\usepackage{graphicx}

\renewcommand{\div}{\nabla \cdot }
\newcommand{\grad}{\nabla}

\newcommand{\Br}{{\bf r}}

\newcommand{\Bu}{{\bf u}}

\newcommand{\Bk}{{\bf k}}
\newcommand{\Bv}{{\bf v}}

\newcommand{\BR}{{\bf R}}

\renewcommand{\BR}{{\bf R}}

\begin{document}

\title{Inverse Scattering and  Acousto-Optic Imaging}
 
\author{Guillaume Bal}
\affiliation{Department of Applied Physics and Applied Mathematics, Columbia University, New York, NY 10027}
\email{gb2030@columbia.edu}

\author{John C. Schotland}
\affiliation{Department of Bioengineering and Graduate Group in Applied Mathematics and Computational Science, University of Pennsylvania, Philadelphia, PA 19104}
\email{schotland@seas.upenn.edu}

\date{\today}

\begin{abstract}
We propose a tomographic method to reconstruct the optical properties of a highly-scattering medium
from incoherent acousto-optic measurements. The method is based on the solution to an inverse
problem for the diffusion equation and makes use of the principle of interior control of boundary measurements by an external wave field. 
\end{abstract}

\maketitle

The acousto-optic effect is a phenomenon in which the optical properties of a material medium are modified after interaction with acoustic radiation. Brillouin scattering from density fluctuations in a fluid~\cite{Born_Wolf} or the ultrasonic modulation of multiply-scattered light in a random medium~\cite{Leutz_1995} are familiar examples of this effect. It is well known that the scattered optical field carries information about the medium. This principle has been exploited to develop a hybrid imaging modality, known as acousto-optic imaging (AOI), which combines the spectroscopic sensitivity of optical methods with the spatial resolution of ultrasonic imaging~\cite{Marks_1993,Kempe_1997,Granot_2001,
Wang_1995,Wang_1998_1,Wang_1998_2,Yao_2000,Li_2002_1,Li_2002_2,Leveque_1999,Leveque_2001,Atlan_2005,
Gross_2009,Lev_2000,Lev_2002}. AOI holds great promise as a tool to probe hidden structure inside of highly-scattering media, such as clouds, paint and biological tissue---a problem which is of both fundamental interest and considerable applied importance~\cite{vanRossum_1999}. For instance, in biomedical applications, optical methods provides unique capabilities to assess physiological function, such as
blood volume and tissue oxygenation~\cite{Ntziachristos_2005, Ripoll_2003, Ripoll_2007}. At the same time, such methods have the molecular selectivity to map gene expression and other markers of bio-molecular activity.
 
In a typical AOI experiment, a highly-scattering medium is illuminated by a coherent optical source and the resulting speckle pattern is registered by a detector. A focused ultrasound beam is then introduced and the speckle modulation is recorded as the beam's focus is scanned throughout the medium. Since the scatterers in the medium are displaced by the acoustic wave, the scattered light undergoes a frequency shift which permits the localization of the resulting so-called tagged photons to the volume containing the focus. The intensity images that are obtained in this manner convey information about the medium. However, they are not tomographic, nor are they quantitatively related to the optical properties of the medium.

In this Letter we consider the inverse scattering problem that arises in AOI. We show that it is possible
to reconstruct tomographic images of the optical properties of a medium of interest from {\em incoherent} measurements of multiply-scattered light that is modulated by a standing acoustic wave. The principle advantages of the proposed method compared to conventional methods for imaging with diffuse light are two fold. (i) The resolution of reconstructed images is, in principle, much higher than in diffuse optical tomography (DOT). In particular, the inverse problem of AOI is well-posed and the image resolution is controlled by the acoustic wavelength. In contrast, the inverse scattering problem for diffuse waves is severely ill-posed, which leads to the relatively low resolution of DOT~\cite{Arridge_1999,Bal_2009}. Physically, the improvement in resolution in AOI can be understood to be a consequence of controlling an internal degree of freedom of the scattering medium (the density of scatterers) by means of an external wave field. (ii) Neither interferometric measurements of tagged photons nor the use of a focused acoustic wave field is required. This considerably simplifies the experimental realization of the method.

We begin by developing a simple model for the acousto-optic effect in a random medium. The model accounts
both for Brillouin scattering and multiple scattering of light due to density fluctuations and is formulated within the framework of radiative transport theory. That is, we do not account for coherent
effects in multiple light scattering~\cite{model}. Consider the propagation of an acoustic wave in a fluid suspension of particles that scatter and absorb light. If the amplitude of the pressure wave is sufficiently small, each particle will oscillate about its local equilibrium position. We may thus regard the suspension as consisting of independent particles whose number density is spatially modulated due to the presence of the acoustic wave. If we focus our attention on a single neutrally buoyant spherical particle, its velocity $\Bu$ obeys the equation of motion
\begin{equation}
\label{eq_motion_u}
\rho \frac{d \Bu}{dt} = \frac{4\pi a \eta}{V} \left(\Bv-\Bu\right) -\grad p \ ,
\end{equation}
where we have not accounted for hydrodynamic interactions between the particles. Here $p$ denotes the pressure, $\Bv$ is the velocity field in the fluid, $a$ is the radius of the particle, $\rho$ is its mass density, $V=4\pi a^3/3$ and $\eta$ is the viscosity of the suspension.
Note that the neglect of hydrodynamic interactions means that, apart from a transient, the particle moves with the fluid.
Consider a standing time-harmonic plane wave of frequency $\omega$ with
\begin{equation}
p = A\cos(\omega t) \cos(\Bk\cdot\Br + \varphi) \ ,
\end{equation}
where $A$ is the amplitude of the wave, $\Bk$ is its wave vector and $\varphi$ is the phase~\cite{wave}.
For simplicity, we have assumed that the speed of sound $c_s$ is constant with $k=\omega/c_s$.
The corresponding velocity field obeys the linearized Euler equation
\begin{equation}
\rho\frac{\partial\Bv}{\partial t} = -\grad p \ .
\end{equation}
 The oscillatory solution to (\ref{eq_motion_u}) is given by
\begin{equation}
\Bu =  \frac{A }{\rho\omega} \sin(\omega t) \sin(\Bk\cdot\Br + \varphi) \Bk \ ,
\end{equation}
and, as a consequence, we obtain for the position of the particle
\begin{equation}
\BR = \BR_0  -  \frac{A}{\rho\omega^2} \cos(\omega t)\sin(\Bk\cdot\Br + \varphi) \Bk\ ,
\end{equation}
where $\BR_0$ is the equilibrium position.

Let $\BR_1, \ldots , \BR_N$ denote the positions of the particles in the fluid. Then, since each particle moves independently, we find that
\begin{equation}
\label{R_i}
\BR_i = \BR_{0i} - \frac{A}{\rho\omega^2} \cos(\omega t)\sin(\Bk\cdot\Br + \varphi) \Bk \ ,   \ \ \ i=1,\ldots, N \ .
\end{equation}
The number density of particles is defined by
\begin{equation}
\varrho = \sum_i \delta(\Br-\BR_i(t)) \ .
\end{equation}
Expanding the delta-function to first order in the small parameter $\epsilon=A\cos(\omega t)/(\rho c_s^2)$ and using~(\ref{R_i}), we find that $\varrho$ is given by
\begin{equation}
\label{number_density}
\varrho(\Br)=\varrho_0(\Br)\left[1 + \epsilon\cos(\Bk\cdot\Br + \varphi)\right] \ ,
\end{equation}
where $\varrho_0(\Br) = \sum_i \delta(\Br-\BR_{0i})$ is the equilibrium number density of the particles.

The propagation of multiply-scattered light is taken to be described by the diffusion approximation to the radiative transport equation. The electromagnetic energy density $u$ obeys the diffusion equation
\begin{eqnarray}
\label{diff_eq}
&&-\div\left[Dn^2\grad \left(\frac{u}{n^2}\right)\right] + c\mu_a u = 0  \quad {\rm in \ \ \ \Omega} \ , \\
\label{bc}
&&u + \ell \frac{\partial u}{\partial n} = \delta(\Br-\Br_0)  \quad {\rm on \ \ \ \partial\Omega} \ ,
\end{eqnarray}
where $n$ is the index of refraction of the fluid, $\mu_a$ is the absorption coefficient, $\ell$ is the extrapolation length and $\Br_0\in\partial\Omega$ is the position of a unit-amplitude point source~\cite{Tualle_2003, Bal_2006}. The diffusion coefficient $D$ is defined by
\begin{equation}
\label{def_D}
D= \frac{c}{3\left[\mu_a + (1-g)\mu_s\right]} \ ,
\end{equation}
where $\mu_s$ is the scattering coefficient and $g$ is the anisotropy of scattering. The scattering and absorption coefficients are related to the number density by $\mu_s=\varrho\sigma_s$
and $\mu_a=\varrho\sigma_a$, where $\sigma_s$ and $\sigma_a$ denote the scattering and absorption cross sections
of the particles. We thus see that the optical properties of the medium are modulated by the presence of the acoustic wave. In addition, we account for variations in the index of refraction due to Brillouin scattering according to
\begin{equation}
\label{ref_index}
n(\Br)=n_0\left[1 + \epsilon\gamma\cos(\Bk\cdot\Br + \varphi)\right] \ ,
\end{equation}
where $n_0$ is the index of refraction in the absence of the acoustic wave and
\begin{equation}
\gamma=\frac{1}{3}\frac{\varepsilon-1}{\varepsilon + 2}
\end{equation}
is the elasto-optical constant, with $\varepsilon$ being the dielectric constant of the fluid~\cite{Born_Wolf}. Note that $\gamma \approx 0.3$ in water. Making use of the above definitions, we see that
(\ref{diff_eq}) becomes
\begin{eqnarray}
\label{diff_eq_psi}
&&-\div D_\epsilon \grad \psi_\epsilon + \alpha_\epsilon \psi_\epsilon = 0 \quad {\rm in \ \ \ \Omega} \ ,  \\
&&\psi_\epsilon + \ell \frac{\partial\psi_\epsilon}{\partial n} =
g \quad {\rm on \ \ \  \partial\Omega} \ ,
\label{bc_psi}
\end{eqnarray}
where $\psi_\epsilon= u/n^2$, $g=n^2 \delta(\Br-\Br_0)$ and the modified absorption and diffusion coefficients are defined by $\alpha_\epsilon=
cn^2\mu_a$ and $D_\epsilon=n^2D$. Using (\ref{number_density}) and (\ref{ref_index}), we find that to first order in $\epsilon$
\begin{eqnarray}
\label{def_alpha_epsilon}
\alpha_\epsilon(\Br) &=& \alpha_0(\Br)\left[1+\epsilon(2\gamma+1)\cos(\Bk\cdot\Br + \varphi)\right] \ , \\
\label{def_D_epsilon}
D_\epsilon(\Br) &=& D_0(\Br)\left[1+\epsilon(2\gamma-1)\cos(\Bk\cdot\Br + \varphi)\right] \ , 
\end{eqnarray}
where $\alpha_0$ and $D_0$ are the absorption and diffusion coefficients in the absence of the acoustic wave. The solution to (\ref{diff_eq_psi}) is given by
\begin{eqnarray}
\label{int_eq}
\nonumber
\psi_\epsilon(\Br) = \psi_0(\Br) - \epsilon\int d^3r' \big[ (2\gamma + 1)G(\Br,\Br')\psi_0(\Br')\alpha_0(\Br') 
\\ + (2\gamma -1)\grad_{\Br'}G(\Br,\Br')\cdot\grad\psi_0(\Br')D_0(\Br')\big]\cos(\Bk\cdot\Br
+ \varphi) + O(\epsilon^2) \ ,
\end{eqnarray}
where the Green's function $G$ satisfies
\begin{equation}
-\div D_0(\Br) \grad G(\Br,\Br') + \alpha_0(\Br) G(\Br,\Br')= \delta(\Br-\Br')
\end{equation}
and obeys homogeneous boundary conditions of the form (\ref{bc}) with zero right hand side. We note that
$\psi_\epsilon$ is proportional to the intensity that is measured by a point detector.

We can estimate the magnitude of the acousto-optic signal from the above analysis. Consider a homogeneous volume of absorption $\alpha$ and diffusion coefficient $D$. It follows from
(\ref{delta_psi}) and (\ref{def_K}) that the relative change in intensity due to the presence of an acoustic plane wave is of the order
\begin{equation}
\frac{\Delta I}{I} \approx \epsilon\left[1+\left(\kappa L\right)^2\right] e^{-\kappa L} \ ,
\end{equation}
where $L$ is the source-detector separation and $\kappa=\sqrt{\alpha/D}$. Choosing typical values of the above parameters in tissue: $\kappa=1
 \ {\rm cm}^{-1}$, $L=1 \ {\rm cm}$ and $\epsilon = 10^{-3}$, we find that $\Delta I/I\approx 10^{-3}$ which is expected to be observable~\cite{numbers}. We note $\Delta I/I$ would be significantly smaller for the case of a focused beam, assuming equivalent incident power.

The inverse problem of AOI is to reconstruct $\alpha_0$ and $D_0$ from $\psi_\epsilon$. To proceed, let us define $\phi=\partial\psi/\partial\epsilon|_{\epsilon=0}$, which can be determined from measurements  carried out in the presence and absence of the acoustic wave.  Making use of (\ref{int_eq}) we obtain
\begin{equation}
\label{delta_psi}
\phi(\Br) = \int K(\Br,\Br') \cos(\Bk\cdot\Br' + \varphi) d^3r' \ ,
\end{equation}
where
\begin{equation}
\label{def_K}
K(\Br,\Br') = \left[(2\gamma+1)G(\Br,\Br')\psi_0(\Br')\alpha_0(\Br')
+ (2\gamma-1)\grad_{\Br'}G(\Br,\Br')\cdot\grad\psi_0(\Br')D_0(\Br') \right] \ .
\end{equation}
Suppose we fix the positions of the optical source and detector and vary the wave vector $\Bk$ and the phase $\varphi$. Evidently, it is then possible to recover $K(\Br,\Br')$ by inversion of a Fourier transform. That is,
\begin{equation}
\label{linear_inversion}
K(\Br,\Br') = \int \frac{d^3k}{(2\pi)^3}e^{-i \Bk\cdot\Br'}\left[\phi(\Br;\Bk,0)+i \phi(\Br;\Bk,3\pi/2)\right] \ ,
\end{equation}
where the dependence of $\phi$ on $\Bk$ and $\varphi$ has been made explicit.
For simplicity, suppose that $\alpha_0=0$. Noting that $G$ and $\psi_0$ depend upon $D_0$,
it follows from (\ref{def_K}) that
\begin{eqnarray}
D_0 &=& \mathcal A[D_0] \ ,
\end{eqnarray}
where the {\em nonlinear} operator $\mathcal A$ is defined by
\begin{equation}
\mathcal \mathcal A[D_0](\Br') =
\frac{K(\Br,\Br')}{(2\gamma-1)\grad_{\Br'} G(\Br,\Br')\cdot\grad\psi_0(\Br')} \ .
\end{equation}
Thus $D_0$ is a fixed point of $\mathcal A$, which can be found iteratively according to
\begin{equation} \label{eq:fixedpoint}
D_0^{(n+1)} = {\mathcal A}[D_0^{(n)}]  \ , \quad n=1,2,\ldots \ ,
\end{equation}
where $D_0^{(n)} \to D_0$ as $n\to\infty$, provided that $\mathcal A$ is contracting. We note that at each step it is necessary to compute the Green's
function $G$, which depends upon the current estimate of $D_0$. We further note that measurements from two independent sources are required to reconstruct both $\alpha_0$ and $D_0$.

The above result provides an iterative solution to the inverse problem of AOI. We now describe a {\em direct} method to solve the inverse problem. We begin by observing that $\phi$ obeys the equation
\begin{eqnarray}
\label{phi}
&&-\div D_0\grad\phi + \left[(2\gamma-1)\div D_0 \grad\psi_0 - (2\gamma + 1)\alpha_0\psi_0\right]\cos(\Bk\cdot\Br+\varphi)
 + \alpha_0\phi = 0 \quad {\rm in \ \  \Omega} \ , \\
&&\phi + \ell \frac{\partial\phi}{\partial n} = 0 \quad {\rm on \ \ \ \partial\Omega} \ ,
\label{bc_phi}
\end{eqnarray}
which follows from (\ref{diff_eq_psi}), (\ref{def_alpha_epsilon}) and (\ref{def_D_epsilon}).
Next, we multiply (\ref{phi}) by $\psi_0$ and (\ref{diff_eq_psi}) by $\phi$, take the difference of the
two equations that result and integrate over $\Omega$. After integrating by parts and employing the boundary conditions (\ref{bc}) and (\ref{bc_psi}), we obtain the identity
\begin{equation}
\Sigma(\Bk,\varphi) =\int_\Omega d^3r \left[(2\gamma - 1)D_0(\grad\psi_0)^2 + (2\gamma + 1)\alpha_0 \psi_0^2  \right]\cos(\Bk\cdot\Br + \varphi) \ ,
\end{equation}
where the surface term $\Sigma$ is defined by the formula
\begin{eqnarray}
\nonumber
\Sigma(\Bk,\varphi)= \frac{1}{\ell}D_0(\Br_0)\left[\phi(\Br_0) +(2\gamma-1)\psi_0(\Br_0)\cos(\Bk\cdot\Br_0 + \varphi)\right] \\
- \frac{(2\gamma-1)}{\ell}\int_{\partial\Omega}d^2r D_0\psi_0^2 \cos(\Bk\cdot\Br + \varphi) \ .
\end{eqnarray}
Since $\phi$ and $\psi$ are known on $\partial\Omega$ from measurements, the function
$\Sigma$ can be determined. Provided that $\Sigma$ is known for a
sufficient number of values of $\Bk$ and $\varphi$, we observe that the
Fourier transform of
\begin{equation}
\label{def_f}
f(\Br) = (2\gamma - 1)D_0(\Br)(\grad\psi_0(\Br))^2 + (2\gamma + 1)\alpha_0(\Br) \psi_0^2(\Br)
\end{equation}
is known from available measurements. We first consider the case where
$\alpha_0=0$. We then find that
\begin{equation}
\label{D_f}
D_0(\Br)= \frac{f(\Br)}{(2\gamma-1)(\grad\psi_0(\Br))^2} \ .
\end{equation}
Making use of (\ref{diff_eq}), we see that $\psi_0$ obeys the nonlinear equation
\begin{eqnarray}
\label{nonlinear_D}
&&\div\left[\frac{f}{(\grad\psi_0)^2}\grad\psi_0\right] = 0 \quad {\rm in \ \ \ \Omega} \ , \\
&&\psi_0 + \ell \frac{\partial\psi_0}{\partial n} = g \quad {\rm on \ \ \ \partial\Omega} \ .
\end{eqnarray}
This equation admits a solution if $f$ is smooth and bounded
from below by a positive constant. Once $\psi_0$ is found by solving (\ref{nonlinear_D}), we
can recover the diffusion coefficient $D_0$ from (\ref{D_f}).

Next we consider the general problem of recovering both $\alpha_0$ and $D_0$. In
this case, we require data from two sources $g_1$ and $g_2$, as specified by
the boundary condition (\ref{bc_psi}). We denote the corresponding solutions to
(\ref{diff_eq_psi}) by $\psi_1$ and $\psi_2$. We also define
\begin{equation}
\label{def_f_12}
f_k(\Br) = (2\gamma - 1)D_0(\Br)(\grad\psi_k(\Br))^2 + (2\gamma + 1)\alpha_0(\Br) \psi_k^2(\Br) \ ,
\quad k=1,2 \ .
\end{equation}
Solving for $\alpha_0$ and $D_0$ we find
\begin{eqnarray}
\label{alpha_coupled}
\alpha_0(\Br) &=& \frac{f_1(\Br)(\grad\psi_2(\Br))^2- f_2(\Br)(\grad\psi_1(\Br))^2}{(2\gamma+1)\left[\psi_1^2(\Br)(\grad\psi_2(\Br))^2-\psi_2^2(\Br)(\grad\psi_1(\Br))^2\right]}  \ , \\
D_0(\Br) &=& \frac{f_2(\Br)\psi_1^2(\Br) - f_1(\Br)\psi_2^2(\Br)}{(2\gamma-1)\left[\psi_1^2(\Br)(\grad\psi_2(\Br))^2-\psi_2^2(\Br)(\grad\psi_1(\Br))^2\right]} \ .
\label{D_coupled}
\end{eqnarray}
The $\psi_k$ are then obtained by solving the system of nonlinear
equations
\begin{eqnarray}
\label{couplednonlin}
&&-\nabla \cdot D_0\nabla \psi_k + \alpha_0 \psi_k = 0
\quad {\rm in \ \ \ \Omega} \ ,  \\
&&\psi_k + \ell \frac{\partial\psi_k}{\partial n} = g_k \quad {\rm on  \ \ \ \partial\Omega} \ ,
\end{eqnarray}
where $\alpha_0$ and $D_0$ are defined by (\ref{alpha_coupled}) and (\ref{D_coupled}), respectively and $k=1,2$. Once the $\psi_k$ are found, we can then recover $\alpha_0$ and
$D_0$ from (\ref{alpha_coupled}) and (\ref{D_coupled}), which yields the solution
to the nonlinear inverse problem of AOI.

Several remarks on the above results are necessary. (i) We observe that the inverse problem of AOI is well-posed. This can be seen from (\ref{linear_inversion}), which yields a bandlimited approximation to the diffusion coefficient by an inverse Fourier transform. In contrast, the linear inverse problem
of DOT is severely ill-posed, requiring the inversion of a Laplace transform, a problem which has logarithmic stability~\cite{Markel_2004,Epstein_2008}. This ill-posedness is responsible for the relatively low resolution of images in DOT. Characterizing the stability of the inverse problem for AOI, along with developing numerical methods for (\ref{couplednonlin}), will be the subject of future research. (ii) The nonlinear equation (\ref{nonlinear_D}) is of the form
\begin{equation}
\label{p-Laplacian}
\div\left(f |\grad \psi|^{p-2}\grad\psi\right) = 0 \ ,
\end{equation}
with $p=0$. The case $p=0$ was also addressed in~\cite{Ammari_2008} in a different physical context and solved numerically using an iterative algorithm similar to the fixed point algorithm presented in (\ref{eq:fixedpoint}). The case $p=1$ was studied in~\cite{Nachman}, where uniqueness results and numerical methods were described. See also related work in~\cite{Kuchment_2009}. (iii) The diffusion equation (\ref{diff_eq}) is valid when the energy density
varies slowly on the scale of the transport mean free path. This condition breaks down when the
acoustic wavelength is sufficiently small. It would thus be of some interest to extend the theory we
have developed to the transport regime.

In conclusion, we have developed a tomographic method for acousto-optic imaging. Neither interferometric measurements of tagged photons nor the use of a focused ultrasound beam is required. Our approach is based on the solution to an inverse problem for the diffusion equation with interior control of boundary measurements.

We would like to thank Shari Moskow for valuable discussions. GB was supported in part by NSF grants DMS-0554097 and DMS-0804696. JCS was supported by the NSF grant DMS-0554100 and the USAFOSR grant FA9550-07-1-0096.


\begin{thebibliography}{99}
\bibitem{wave}
We do not consider directly how such a wave is to be generated. Evidently, 
an appropriately chosen boundary condition for the acoustic wave equation will, 
in principle, solve the problem.
\bibitem{numbers}
To estimate $\epsilon$, we chose $\rho=1 \ {\rm g \ cm^{-3}}$, $c_s=1.5\times 10^5 \ {\rm cm \ s^{-1}}$ and $A=10^6 \ {\rm Pa}$.
\end{thebibliography}
\end{document}